# Risk-Free Rate in the Covid-19 Pandemic:
# Application Mistakes and Conclusions for Traders

Magomet Yandiev

The Faculty of Economics, Lomonosov Moscow State University

mag2097@mail.ru

## 1. Introduction

This short paper is intended to demonstrate a crucial omission made by traders in setting the risk-free interest rate, especially in times of crisis: instead of increasing the risk-free rate, traders undercut it en masse on the contrary. This results in incorrect investment and financial decisions, especially those involving CAPM models, option pricing models and portfolio theory.

## 2. Literature Review

Literature on risk-free rate setting was searched on the international library's Social Science Research Network website, www.ssrn.com/, using the keywords "risk-free rate" in the title, abstract and keywords of the publications in the library[1]. 1,644 papers were retrieved; search using the same keywords, but in the title only, yielded 50 results, of which 30 papers relevant to the subject of this paper were selected (see 7. References). In doing so:

- papers with just over one hundred downloads were selected irrespective of the year they were published in; for reference, the earliest paper dates back to 1995;

- papers published in the last five years were selected without regard to the number of downloads, i.e. even if the number of downloads was less than one hundred but more than one;

- papers published in the last five years but never downloaded were ignored.

Based on the year of each selected publication, Graph 1 was plotted. It shows that interest in the risk-free rates has grown in the last decade as compared to the noughties and the last century. However, the publishing activity is low as compared to other financial subjects, although the risk-free rate underpins some of the most important financial models.

---

[1] The web search engine searches by individual words not whole phrases

**Graph 1. Number of publications on the subject**

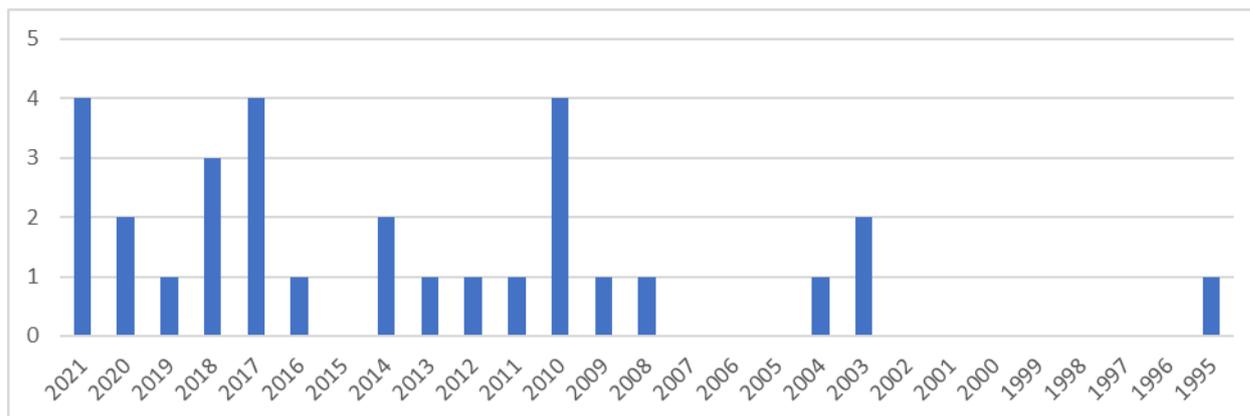

*Source*: references, author's calculations

It is conceivable that the financial crises of recent decades have somewhat intensified the scientific inquiry, in particular the role and impact of the risk-free interest rate on financial and economic parameters. The papers included in the list of references cover a wide range of areas, from macroeconomics to real estate. For example, it is argued that the inflation variability in the economic policy leads to reduction in the risk-free rate [28], and when the risk-free rate decreases, inflows into the US mutual funds become less sensitive to past returns [8]. Other papers explain the negative relation between corporate bond spreads and risk-free interest rates [24] and state that the treasuries interest rate is lower than the risk-free interest rate [16] – these calls into question the choice of risk-free interest rate based on the traditional rule that the lower the bond yield, the lower the risk. Paper [13] demonstrates that the lower the risk-free rate, the higher the demand for risky assets and [30] states that the correlation of share prices and risk-free rate is insignificant, and it sounds as if the risk-free rate changes as often as the share price. It is argued that the risk-free rate contains economic considerations that affect dividends [26] and that changes in the risk-free rate are a decisive factor in changes in the housing prices [21] - the latter demonstrates the cross-sectoral importance of the risk-free rate.

In an effort to expand the concept of the risk-free rate, to improve the accuracy of its application, and possibly facing the unexplained phenomena in the markets, scholars have introduced various interest rates, not always justified in my opinion. For instance, paper [2] analyses the "fixed" and "variable" risk-free interest rates, [9] the "implied", [11] the "normalised", [15] the "upper" and "lower" limits of the risk-free interest rate, [20] the "equilibrium", [21] the "real", [28] the symbiosis of the latter two - the "equilibrium real" and [25] the "global". Paper [18] introduces the term "volatility of the risk-free rate", which defies all logic: a risk-free asset cannot be volatile, while paper [1] uses the term "random changes" of the risk-free rate, which is also proved controversial, because the risk-free rate is determined not by the market to be able to "randomly change", but by the traders themselves, and besides, it seems to be a rather stable parameter.

Some papers emphasise that scholars and practitioners disagree about what appropriate source is to be chosen to set the risk-free rate [5], the source to determine the risk-free rate and even the concept of the risk-free rate are criticised [6]. These issues are apparently so relevant that they give rise to a deadly question: "what if there are no risk-free investments" [7]. The risk-free criterion is the absence of bond defaults. It is difficult for us to accept this approach, because there is no issuer, whether governmental or corporate, that would issue bonds allowing for its default (apart from a malicious intent). But then, most securities traded on the bond market could be considered risk-free.

Numerous different recommendations regarding the risk-free rate can be found in the academic literature. For example, paper [6] questions the choice of government returns as a source of risk-free returns and argues that exchange rates could be a good alternative to the risk-free rate. In [14], the risk-free interest rate is defined as the return on an infinitely diversified portfolio, but by no means as a parameter determined by expert methods. Paper [10] draws on the use of alternative risk-free rates with a focus on LIBOR. In [27] the calculation of risk-free rates for money instruments is also discussed, paying particular attention to situations where the maturity of money instruments does not coincide with the maturity of other assets studied. It is asserted in [22] that treasury bills reflect the risk-free rate better than long-term treasury securities, regardless of the investment horizon. It is acknowledged in [22] that scholars and practitioners tend to use short-term treasury bills or long-term treasury bonds as risk-free securities without empirical support.

Thus, the academic literature does not resolve the scientific uncertainties with respect to the risk-free rate issue.

## 3. The Whole Truth of the Risk-Free Rate

An interesting feature of the risk-free interest rate[2] is that all financiers know of it, recognise its necessity, but apart from saying that it is a rate of a risk-free instrument return, there is little else they can add: there is an investment with guaranteed return that offers both firms and investors a "risk-free" option [7].

Nevertheless, let us take what has been said and expand it a little: the risk-free rate of return is the rate of return on an asset the standard deviation of which is zero. It follows that a risk-free instrument is any financial asset with zero volatility.

---

[2] Short aside: science covers the subject studied in detail. In our study of the risk-free interest rate, it is logical to check whether it would be reasonable to introduce and apply the opposite term, "zero-yielding risk", and determine the relation between such terms. However, we have found no paper on this subject in the library. In practice, of course, all traders attempt to avoid taking a risk without reward, otherwise they will incur losses. And no matter what attempts they make, that is frequently the case in practice.

.

Based on this, the following known sources used to determine the risk-free rate can be listed and new ones can be formulated.

**1. Government bonds**. A risk-free rate is most commonly chosen when using the return to maturity of government bonds. It is a low-cost, low-conflict and quick choice of the rate.

**2. Bank deposits.** Using the bank deposit rate is also popular because of its simplicity.

**3. Short-term loans on the interbank loan (IBL) market.** LIBOR can be used as a risk-free rate, but it is very controversial because the lending period in the interbank market is very short and does not allow the use of this rate for medium- and long term transactions.

**4. Constructor.** This is a calculation of the risk-free rate based on beliefs such as "The asset return is a sum of the risk-free return and the risk premium for investing in securities of that issuer" or "The risk-free return in Russia is a sum of the return on the US treasury securities plus the risk premium for investing in Russia" and similar. The risk-free rate thus obtained will be purely of an expert nature.

However, this list is not exhaustive. Two more sources can be added.

**5. Zero Beta shares.** Purely theoretically, there can be such share in the market whose Beta factor is zero or as close to zero as possible. According to the CAPM model, the expected return on an asset with zero Beta is equal to the risk-free rate. Moreover, such share can safely be considered risk-free because it is not volatile. Hence, a share with zero Beta is another source to determine the risk-free rate. It sounds almost fantastic, but nothing prevents us from using such shares in practice and they bring their owners a truly risk-free income.

**6. Arbitrage deals.** Arbitrage deals, popular on financial markets, are absolutely unfairly disregarded. Most speculators seek arbitrage opportunities and convert them into profit. There is no formal risk in such deals. A simple example: there are two exchanges, on which shares of the company are traded; on the first exchange they cost twice as much as on the second; then the speculator borrows shares and sells them on the first exchange and uses the proceeds to buy the same shares on the second exchange; the result: debt in shares, available shares to cover the debt and profit. Of course, arbitrage deals are not a financial asset, they are a series of deals with different assets, but in general the whole set of arbitrage manipulations looks, in the end, exactly like investing in one notional asset. Therefore, returns on arbitrage deals can rightly be regarded as risk-free.

Thus, we have six sources of setting a risk-free rate. Logically, out of the six options, the asset with the highest return should be chosen as a risk-free option. This is logical, because no one would choose an asset with a low return if there is an equal risk-free asset but with a higher return.

However, this thesis should be rejected because financial institutions can operate simultaneously in different segments of financial markets and may have priorities in favour of one or another market. If any company conducts deals across segments evenly, then the arithmetic average of all six sources can be taken as a risk-free average. However, this is not the case in real life and different companies invest different volumes in different segments of financial markets. Therefore, the risk-free interest rate should be taken as a weighted average of all six sources for the volume of investment in a particular segment of the financial markets.

All of the above can be summarised as follows:

$$r0 = W1*r1 + W2*r2 + W3*r3 + W4*r4 + W5*r5 + W6*r6 \qquad [1]$$

where:

r0 is a risk-free rate;

W1, r1 is a share of funds invested in bonds in the total volume of deals; bond yield to maturity;

W2, r2 is a share of funds invested in bank deposits in the total volume of deals; return on bank deposits;

W3, r3 is a share of funds invested in loans in the interbank loans market in the total volume of deals; return on interbank loans;

W4, r4 - the share and return are determined expertly, which entails, *inter alia*, adjusting expertly all the other parameters (W1-4, W6) downward;

W5, r5 is a share of funds invested in zero Beta shares in the total volume of deals; current return on such shares;

W6, r6 is a share of funds invested in arbitration in the total volume of deals; return on arbitrage deals.

Thus, each trader will have its own risk-free rate. It reflects the company's opportunities to generate a risk-free income. Some companies have more opportunities, and others have less, so each trader should use its own risk-free rate, and avoid focusing on certain general values, for example, the average national value.

Formula 1 shows that the current understanding of the risk-free interest rate (first four factors only) is incomplete and, therefore, the value of the risk-free rate actually used in practice is incorrect.

## 4. Risk-Free Interest Rate in Crisis

Incomplete understanding of the risk-free rate is clearly manifested in crises, for example, during the ongoing Covid-19 pandemic, we see a global decline in the risk-free interest rate [12].

Of course, the following weaknesses of the study should be noted [12]:

- the risk-free rate is calculated by all traders in different ways, there is no single international method, therefore, it is difficult to benchmark;

- the average country-wide rate is analysed, although the spread of values in each country is quite high; the largest range is demonstrated by Argentina: the risk-free rate used by local traders ranges from 5.5% to 47.8%.

Despite the criticism, the study [12] is practically the only academic source of information on the risk-free interest rate in the world. So, according to this study:

- 2021 vs. 2020: in most mainly developed countries (22 countries), the risk-free rate has decreased in 2021 vs. 2020;

- 2020 vs. 2019: the vast majority of countries (34 countries) saw a decrease in the risk-free rate in 2020 vs. 2019;

- 2019 vs. 2018: most countries (25 countries) experienced a decrease in the risk-free rate in 2019 vs. 2018;

- 2018 vs. 2015: 26 countries recorded an increase in the risk-free rate in 2018 vs. 2015.

So, the paper [12] shows that the risk-free rate has been mainly decreasing in the last three years. This contradicts Formula 1.

To prove the above, we will firstly analyse Formula 1 and for better illustration of the main idea will expertly establish the criteria to define weights of all components in the formula in relative sizes (see Table 1).

**Table 1. Weight of Components in Formula 1**

| Formula 1 components | W1*r1 | W2*r2 | W3*r3 | W4*r4 | W5*r5 | W6*r6 |
|---|---|---|---|---|---|---|
| Component contents | bonds | bank deposits | IBL loans | expertly | zero Beta | arbitrage opportunities |
| Significance of components in normal conditions | high | average | low | average | minimum | minimum |
| Significance of components in crisis | low | low | minimum | minimum | minimum | high |

*Source*: prepared by the author

The specified criteria allow us to determine approximate changes in the weights of the formula after the onset of crisis. Reasoning for proof:

- in crisis, some issuers incur losses and others earn additional profits;

- this fact forces traders to reconsider the current quotations offered by them in trading, portfolios, etc.;

- collective reinterpretation of quotations results in growing asset volatility;

- increased volatility creates many new arbitrage opportunities in financial markets that were previously impossible;

- traders increase their investments and even concentrate their resources in arbitrage deals.

Consequently, after the onset of crisis, the arbitrage deal factor in formula 1 will increase significantly and all other factors will no longer be important, which means that the risk-free rate will eventually grow.

Thus, accumulation of crisis phenomena should be accompanied by an increase in the weight of the risk-free interest rate, rather than a decrease as is the case with most countries now. Decrease in the risk-free rate has a negative impact on investment and financial decision-making, sets wrong benchmarks for business development and causes mistakes in the application of the risk-free rate.

## 5. Consequences of mistakes in applying the risk-free rate

### 5.1. In the CAPM model

Reducing the risk-free rate in times of crisis, as practised around the world and disputed by the author, causes the return on an asset determined by the CAPM model (with all other parameters remaining constant) to increase; although it should decrease. Under such conditions, the traders are forced to sell an asset rather than buy it, and vice versa. Consequently, the percentage of erroneous decisions in traders' activities increases.

### 5.2. In portfolio theory

As we know, the portfolio theory recommends the inclusion of a risk-free asset in the portfolio in order to simplify the optimal portfolio determination. However, reducing the risk-free rate in times of crisis, as is practiced around the world and disputed by the author, leads to the traders incorrectly setting a weight of the risk-free instrument in the asset portfolio. The key point here is that the risk-free rate in crisis should actually increase and at some point the risk-free instrument line will become almost horizontal and, with an even larger increase in the risk-free asset, it will incline in the opposite direction (see Figure 1). But then a risk-free asset included in the securities portfolio will cease to be useful: the trader will be no longer able to control the portfolio risk and return; it will be a burden to the latter.

**Figure 1:** New understanding of the risk-free rate in the portfolio theory in crisis.

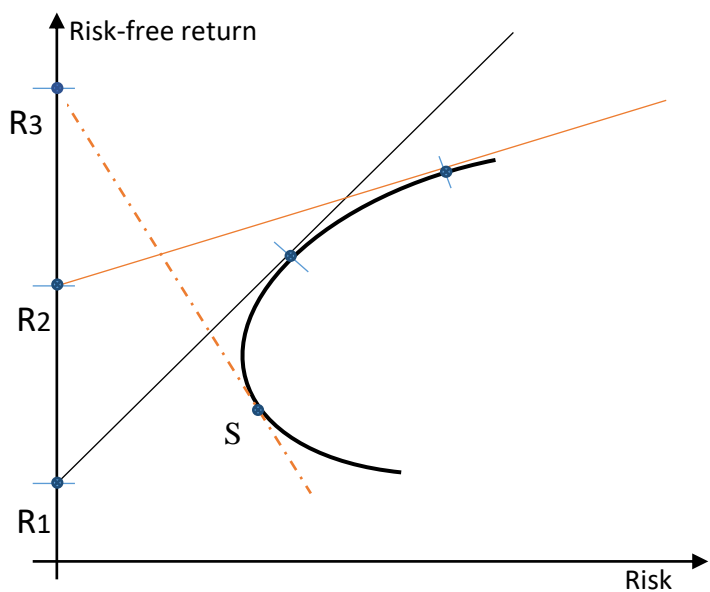

*Source*: developed by the author

Figure 1 shows the line starting at R1, illustrating traditionally the application of the risk-free rate in the portfolio theory. When the value of the risk-free rate (point R2) increases, the slope of the tangent to the set of portfolios becomes flatter. A small increase in profitability is accompanied by a significant increase in risks. But the most interesting thing is observed when the risk-free rate rises significantly (point R3): the slope of the line changes and the tangent can only be drawn to the bottom of the graph, offering an obviously unprofitable solution, as higher returns can be obtained at the same level of risk. There is a clear inefficiency.

5.3. In the option pricing model

With reduction in the risk-free rate in times of crisis, as is practiced around the world and disputed by the author, the Black-Scholes option pricing model, all things being equal, shows a price decrease, whereas the opposite should be the case. That is why mistakes are made in financial decisions.

**6. Conclusions**

1. Formula 1 shows that the current understanding of the risk-free rate (first four factors from formula 1 only) is incomplete and that is why the risk-free rate used by traders is incorrect.
2. An increase in economic crisis phenomena should be accompanied by the growth in the risk-free rate.

3. The growth of the risk-free rate is somewhat limited. Once this limit has been reached, the application of the risk-free instrument to establish a securities portfolio will become inexpedient.
4. The risk-free rate can be set on the basis of the return of shares with zero Beta factor.
5. The risk-free rate can be set on the basis of the return on arbitrage deals.